\documentclass{PoS}
\usepackage{enumitem}   
\usepackage{braket}
\usepackage[utf8]{inputenc}
\usepackage{import}
\usepackage{braket}
\usepackage{multirow}
\usepackage{here}
\usepackage{amsmath}
\usepackage{subfigure}

\title{Meson interactions at large $N_c$ from Lattice QCD}

\ShortTitle{Meson interactions at large $N_c$ from Lattice QCD}

\author{\speaker{Fernando Romero-L\'opez}\\
        IFIC, CSIC-Universitat de Val\`encia, \\ 46980 Paterna, Spain \\
     \email{fernando.romero@uv.es}}
 \author{Andrea Donini \\  IFIC, CSIC-Universitat de Val\`encia,\\ 46980 Paterna, Spain \\
     \email{donini@ific.uv.es}
        }

        \author{Pilar Hern\'andez \\  IFIC, CSIC-Universitat de Val\`encia,\\ 46980 Paterna, Spain \\
     \email{m.pilar.hernandez@uv.es}
        }
        \author{Carlos Pena \\  Departamento de F\'isica Te\'orica and IFT UAM-CSIC, Universidad Aut\'onoma de Madrid, \\ 28049 Madrid, Spain \\
     \email{carlos.pena@uam.es}
        }
      
\abstract{We report on the computation of the scaling of QCD observables with the number of colours, $N_c$. For this, we use dynamical configurations with four active flavours, $N_f=4$, and values of $N_c=3-6$. We study the meson masses and decay constants, and compute the leading and subleading contributions to the Low Energy Constants (LECs) of the chiral Lagrangian. We also explore $\pi \pi$ scattering in the $I=2$ channel, and compute the $K \to \pi $ weak decay matrix elements. We comment on the relation of the latter to $K \to \pi\pi$ processes  and the $\Delta I=1/2$ rule.}

\FullConference{37th International Symposium on Lattice Field Theory - Lattice2019\\
		16-22 June 2019\\
		Wuhan, China}

\begin{document}

\section{Introduction}

The 't Hooft limit of QCD  \cite{tHooft:1973alw}, that is the limit of infinite number of colors, $N_c$, constitutes a simplification of the theory of the strong interactions that preserves most of its non-trivial properties, such as confinement or spontaneous chiral symmetry breaking. The large $N_c$ limit of QCD is a powerful approximation. For instance, it has been able to predict the hierarchy of the low energy constants (LECs) of the chiral Lagrangian. Therefore, many other phenomenological approaches to QCD often use approximations inspired by the large $N_c$ limit. Still, there are some exceptions where large $N_c$ fails. The most striking being the ratio of isospin amplitudes in the $K \to (\pi \pi)_I$ weak decay, in which the two pions can be in an isoscalar ($I=0$) or isotensor ($I=2$) final state. This ratio, $A_0/A_2$, is measured to be $\sim 22$, although large $N_c$ predicts it to be $\sqrt{2}$. This is the so-called ``$\Delta I = 1/2$ puzzle'', whose solution remains elusive. 

Lattice Field Theory (LFT) offers the possibility of \textit{ab initio} explorations of generic gauge theories, with arbitrary numbers of colours and flavours --- see Ref. \cite{marga} for a review and Refs. \cite{Bali:2013kia,DeGrand:2016pur} for benchmark calculations. In this context, simulating QCD at a different values of $N_c$ has a considerable amount of yet unexplored potential. In particular, the study of the $N_c$-scaling of QCD observables can help to improve phenomenological predictions. An illustrative example can be found in Ref. \cite{Gisbert:2017vvj}, where the systematic error induced by the large $N_c$ limit completely dominates the total uncertainty. In addition, a nonpertubative study of the $N_c$-scaling of kaon decay amplitudes may reveal hidden features of QCD at the origin of the $\Delta I=1/2$ rule. \footnote{Lattice QCD attemps have observed the experimental enhancement although with large error bars  \cite{Bai:2015nea,Blum:2015ywa, Ishizuka:2018qbn}.}

Our project consists in the computation of physical observables at different values of $N_c$ \cite{Donini:2016lwz,Romero-Lopez:2018rzy,Hernandez:2019qed}. Specifically, we are currently focusing on meson masses and decay constants, kaon decays and meson-meson scattering.  In this talk, we aim at summarizing the current status of the project and present several results --- some of which are preliminary --- for the aforementioned observables. 

\section{Simulations at large $N_c$}

We are using configurations with $N_f=4$ and $N_c=3-6$ generated with HiRep \cite{DelDebbio:2009fd}. We use the Iwasaki gauge action, and $O(a)$-improved Wilson fermions. A summary of the values of the simulation parameters can be found in Ref. \cite{Hernandez:2019qed}. We use the gradient flow scale $t_0$ (see Ref. \cite{Luscher:2010iy}) for the scale setting \cite{Romero-Lopez:2018rzy,Hernandez:2019qed}. This way, the lattice spacing is kept constant across all values of $N_c$, and it is $a \sim 0.075$ fm. Furthermore, we use a mixed-action setup, that is, twisted mass fermions at maximal twist in the valence sector. This has several advantages:
\begin{enumerate}[label=(\roman*)]
\item The computation of the meson decay constant requires no renormalization constant. It can be obtained from the bare pseudoscalar correlation function and the value of the bare twisted mass $\mu_0$, via the PCAC relation:
\begin{equation}
C(t) = \braket{P(t) P(0)} \xrightarrow{\ \ \ T \gg t \gg 1 \ \ \ } |  \braket{0 | P | \pi}_{0}   |^2 e^{-M_\pi t}, \ \ \ \ \ \ \ \ \  F_\pi = \frac{2 \mu_0  \braket{0 | P | \pi}_{0} }{M_\pi^2}.
\end{equation}
\item The evaluation of weak matrix elements using twisted mass fermions at maximal twist helps alleviating renormalization problems, e.g., mixing with wrong-chirality operators (see Ref. \cite{Frezzotti:2004wz}). 
\item Several observables of interest with twisted mass in the valence sector show smaller statistical fluctuations. In addition, the twisted mass formulation prevents the appearance of linear cutoff effects in $F_\pi$ and in the weak matrix elements.
\end{enumerate}

\section{Large $N_c$ scaling of meson masses and decay constants \label{sec:LECs}}

\begin{figure}[h!]
   \centering
   \subfigure[Meson decay constant LECs]%
             {\includegraphics[width=0.5\textwidth,clip]{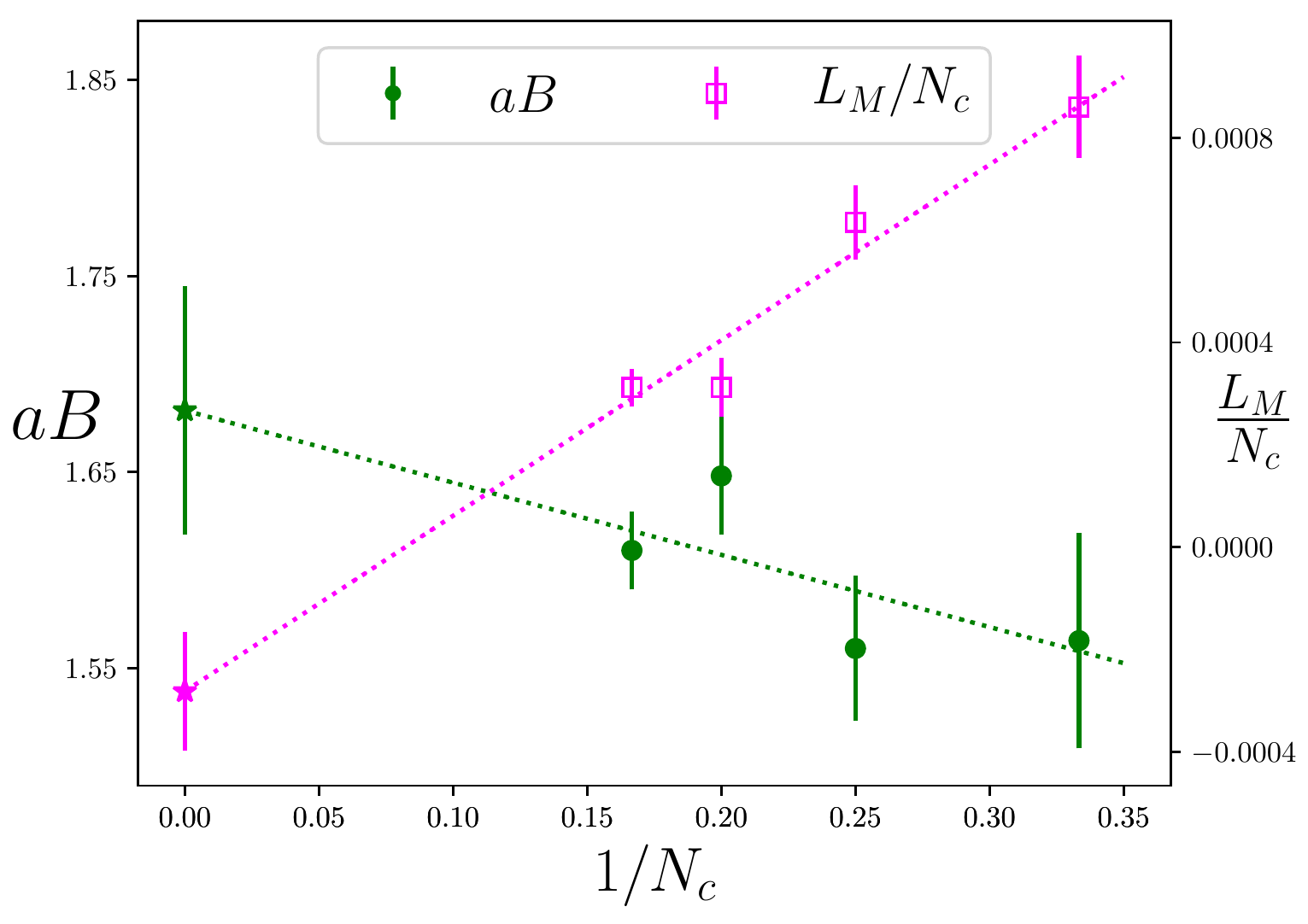}}\hfill
   \subfigure[Meson mass LECs]%
             {\includegraphics[width=0.5\textwidth,clip]{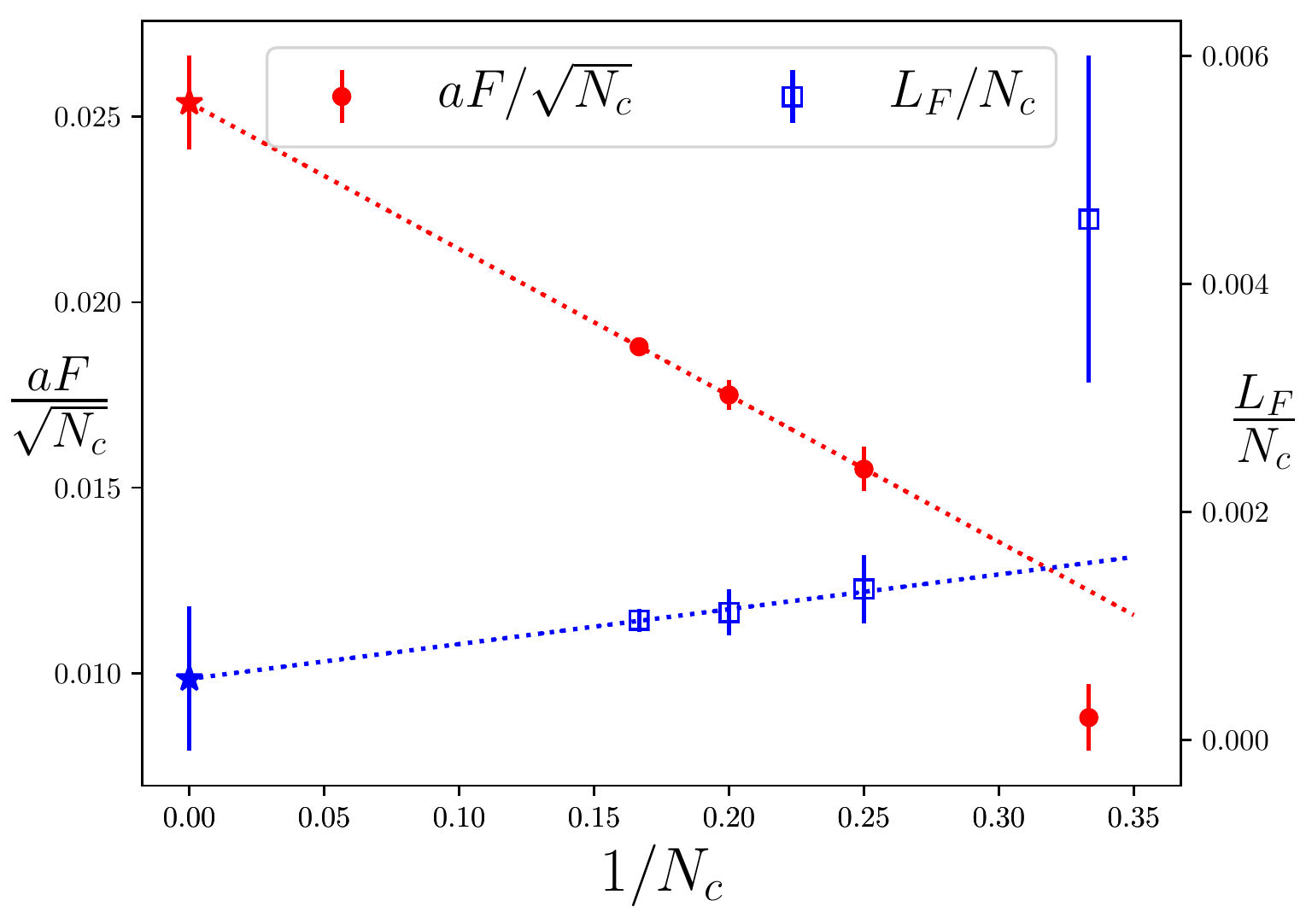}}
   \caption{$N_c$-scaling of the LECs of the chiral Lagrangian with the number of colors for the non-singlet pseudoscalar meson decay constant (left) and mass (right). }
   \label{fig:LECs}
\end{figure}

The computation of the large $N_c$ scaling of meson masses and decay constants was recently published in Ref. \cite{Hernandez:2019qed}. In this section, we aim to summarize the approach and highlight some relevant results.  We start by pointing out that the chiral perturbation theory (ChPT) description of pseudoscalar mesons is valid for an arbitrary number of colors \footnote{Two-coloured QCD is the exception as mesons and baryons have the same quark content, see Ref. \cite{Drach:2017btk} for recent calculations in this theory. }. The prediction of next-to-leading order (NLO) ChPT with $N_f$ degenerate flavours for the non-singlet meson mass and decay constant are \cite{Bijnens:2009qm}:
\begin{align}
\begin{split}
F_\pi = {F} \Bigg[1   - {\frac{N_f}{2} \frac{M_\pi^2}{(4 \pi F_\pi)^2}\log \frac{M_\pi^2}{\mu^2}}     +4 \frac{M_\pi^2}{F_\pi^2} \Big( L_5^r  +  N_f L_4^r  \Big)  \Bigg], 
\label{eq:Fpi1}
\end{split} \\
\begin{split}
M^2_\pi = 2B m \Bigg[1  + {\frac{1}{N_f} \frac{M_\pi^2}{(4 \pi F_\pi)^2}\log \frac{M_\pi^2}{\mu^2}}  +8\frac{M_\pi^2}{F_\pi^2} \Big( N_f (2 L_6^r  - L_4^r ) + 2L_8^r - L_5^r  \Big)  \Bigg], \label{eq:Mpi1}
\end{split}
\end{align}
in terms of the leading order (LO) couplings, $B, F$, and the NLO Gasser-Leutwyler coefficients,  $L^r_{4,5,6,8}(\mu)$, defined at the renormalization scale $\mu$. 
Even though there is no explicit $N_c$ dependence in Eqs.~(\ref{eq:Fpi1}-\ref{eq:Mpi1}), the LECs scale with $N_c$ as:
\begin{eqnarray}
\begin{array}{ll}
O(N_c): F^2, L_5, L_8;&  \ \
 O(1): B, L_4, L_6.
 \end{array}
 \label{eq:nc}
\end{eqnarray}
Loop corrections are suppressed in $1/F_\pi^2 = O(1/N_c)$, and hence the loop expansion is expected to converge better at larger $N_c$. Keeping only leading and subleading dependence on $N_c$ for the LECs, convenient parametrizations are:
\begin{align}
& \ \ \ \ \ \ \ \ \ \ \ \ \ \ \ \ \ \ \ \ \ \ \ \ \ \ \ \ \ \ \ F = \sqrt{N_c} \left( {F}_0 + \frac{{F_1}}{N_c}  \right), \ \ \ \ \ B =   {B}_0 + \frac{{B_1}}{N_c},  \label{eq:lonc} \\
 &L_5  + N_f L_4 \equiv L_F = N_c L_F^{(0)}  + L_F^{(1)},  \ \ \ \ \
  2 L_8 - L_5 + N_f (2 L_6 - L_4 )\equiv L_M = N_c L_M^{(0)}  + L_M^{(1)}.
\label{eq:lecsnc}
\end{align}
Note that according to the scaling of Eq.~(\ref{eq:nc}) and the definition of Eq. (\ref{eq:lecsnc}):
\begin{align}
\begin{split}
&L_F^{(0)} = {L_5 \over N_c} + {\mathcal O}\left({1 \over N_c}\right), \ \ \ \ \ L_M^{(0)} = {2 L_8 - L_5 \over N_c} + {\mathcal O}\left({1 \over N_c}\right).
\end{split}
\end{align}
We have performed fits to the expressions in Eqs. (\ref{eq:Mpi1}) and (\ref{eq:Fpi1}) at fixed values of $N_c$. The results are presented in Fig. \ref{fig:LECs}.  As can be seen, Eqs. (\ref{eq:lonc}) and (\ref{eq:lecsnc}) seem to describe the scaling properly, with the exception of the decay constant at $N_c=3$, where subleading effects seem larger.

Once we have checked the ansätze from Eqs.  (\ref{eq:lonc}) and (\ref{eq:lecsnc},) it is better to perform a simultaneous chiral and $N_c$ fit. In this fit, the number of parameters is small in comparison to the number of degrees of freedom. For the decay constant, this yields:
\begin{align}
\begin{split}
& \frac{F}{\sqrt{N_c}} = \left( 67(3) - 26(4) \frac{N_f}{N_c}\right)\  (3\%)^a \ \ \text{ MeV}, \
 \ \ \ \ \ \ \frac{L_F(\mu)}{N_c} \cdot 10^3 = 0.1(4) + 0.6(3) \frac{N_f}{N_c},  \label{eq:fitsFpi}
\end{split}
\end{align}
with a $3\%$ of scale setting error, and where we have assumed the $N_f$ dependence, as argued in Ref. \cite{Hernandez:2019qed}. From the previous fit results, we can infer results for QCD with $N_f=3$:
\begin{equation}
F^{N_c=3,N_f=3} = 71(3)\text{ MeV }, \ \ \ \ L_M^{N_c=3,N_f=3} = 2.1(3) \cdot 10^{-3},
\end{equation}
which are in agreement with other phenomenological and lattice determinations --- See Ref. \cite{Aoki:2019cca} for a summary of current results. 

We conclude this section by mentioning that in the large $N_c$ limit, the flavour singlet ($\eta'$) becomes a Goldstone boson. Thus, one should also repeat the analysis using $U(N_f)$ ChPT. We refer the reader to our publication for more details \cite{Hernandez:2019qed}. In addition, the mass and decay constants of the flavour singlet are currently under exploration.

\section{I=2 $\pi \pi$ scattering in the large $N_c$ limit}

As mentioned above, the large $N_c$ limit preserves quark confinement and chiral symmetry breaking. However, the interactions between mesons become weaker as $N_c$ increases. Thus, QCD at large $N_c$ is made up of infinitely narrow non-interacting resonances. The goal of this section is the study of the least complicated two-hadron interaction in the large $N_c$ limit: $\pi\pi$ scattering at maximal isospin, $I=2$.

Within lattice QCD, the study of two-particle interactions is perfomed using the Lüscher method \cite{Luscher:1986pf}. It is an indirect way of obtaining two-particle scattering amplitudes from the spectrum of a theory in finite volume. In its simplest form, the Lüscher formalism relates the $s$-wave scattering length, $a_0$, of two identical (pseudo)scalar particles to the energy shift of the two-particle ground state in finite volume. This expression, valid only for $L \gg a_0$, is called the threshold expansion and takes the form:
\begin{equation}
\Delta E  = E_{\pi\pi} - 2 M_\pi = -\frac{4\pi a_0}{m L^3} \left[ 1+ c_1 \frac{a_0}{L}  + c_2 \left( \frac{a_0}{L} \right)^2   \right] + O(L^{-6}),
\end{equation}
with $c_1,c_2$ numerical constants (see Ref. \cite{Luscher:1986pf}).

The low-energy pion interactions are in general well-described by ChPT. The leading order prediction for the $I=2$ channel is given by:
\begin{equation}
M_\pi a_0^{I=2} = - \frac{M_\pi^2}{16 \pi F_\pi^2} + O\left[\frac{M_\pi^4}{(4 \pi F_\pi)^4}\right],
\end{equation}
and thus, by knowing that $F_\pi^2 \sim O({N_c})$ and $M_\pi \sim O(1)$, we can easily extract the leading $N_c$-scaling:
\begin{equation}
M_\pi a_0 \propto \frac{1}{N_c} + O(N_c^{-2}).
\end{equation} 

\begin{figure}[h!]
\centering 
\includegraphics[width=.7\textwidth]{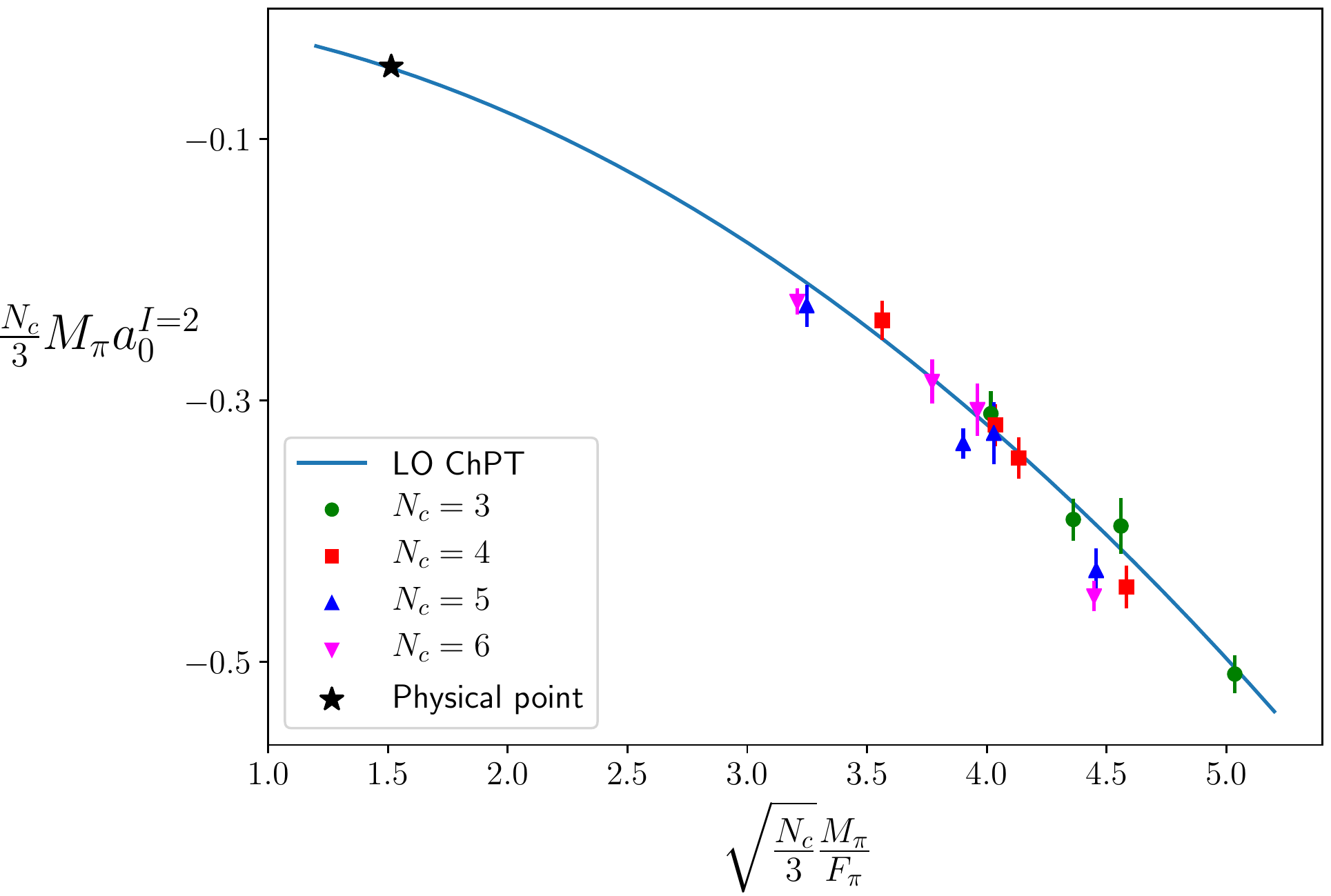}
\caption{\label{fig:a0} Preliminary results on the $s$-wave I=2 $\pi\pi$ scattering length for different values of $N_c$. We rescale the $x$- and $y$-axis by the leading $N_c$-dependence. The line indicates the prediction of leading order chiral perturbation theory, and the black star marks the value for physical pion masses.}
\end{figure}

The results for $a_0^{I=2}$ in our 16 ensembles can be found  in Fig. \ref{fig:a0}. We display our results by multiplying both the $x$- and $y$-axis with the leading $N_c$ dependence, so that points at different values of $N_c$ should lie on a universal line up to subleading $1/N_c$ corrections. As it can be seen, our results show excellent agreement with leading order ChPT even at pion masses of $M_\pi = 560 $ MeV, corresponding to the heaviest point in Fig. \ref{fig:a0}.

The NLO prediction of ChPT for this scattering channel exists and includes the NLO LECs of the chiral Lagrangian. In a next step of our work, the data of Fig. \ref{fig:a0} will be used to try to extract the $N_c$-scaling of the LECs in a similar manner to the one explained in Section \ref{sec:LECs}.

\section{Nonleptonic kaon decays at large $N_c$}

The study of the $\Delta I = 1/2$ rule in the large $N_c$ limit can be performed via an indirect method: the study of the $K \to \pi$ matrix elements in the GIM limit \cite{Donini:2016lwz,Giusti:2006mh}. We refer to our previous work for technical details \cite{Donini:2016lwz,Romero-Lopez:2018rzy}. The key point is that the ratio of isospin amplitudes in the GIM limit can be related to leading order in ChPT to the $K \to \pi$ matrix elements:
\begin{equation}
\frac{A_0}{A_2} = \frac{1}{2\sqrt{2}} \left(1 + 3 \frac{g^-}{g^+} \right), \ \ \ \ \ \ \braket{K | C_\pm \mathcal{O}^{\pm} | \pi } = g^\pm, \label{eq:ratiog}
\end{equation}
where $\mathcal{O}^{\pm}$ are the operators in the effective weak Hamiltonian, $C_\pm$ are the Wilson coefficients and  $g^\pm$ are the LECs of the chiral weak Hamiltonian.

On the lattice, the effective couplings $g^\pm$ can be determined via a three-point function properly normalized:
\begin{eqnarray}
g^\pm \sim {R}^\pm \equiv \frac{\langle\pi|\mathcal{O}^\pm|K\rangle}{f_Kf_\pi m_Km_\pi}
= Z_R^\pm(\mu) R^\pm_{bare} \,,
\label{eq:ratio}
\end{eqnarray}
and the Wilson coefficients can be calculated as explained in Ref. \cite{Donini:2016lwz}. 

\begin{figure}[h!]
\centering 
\includegraphics[width=0.6\textwidth]{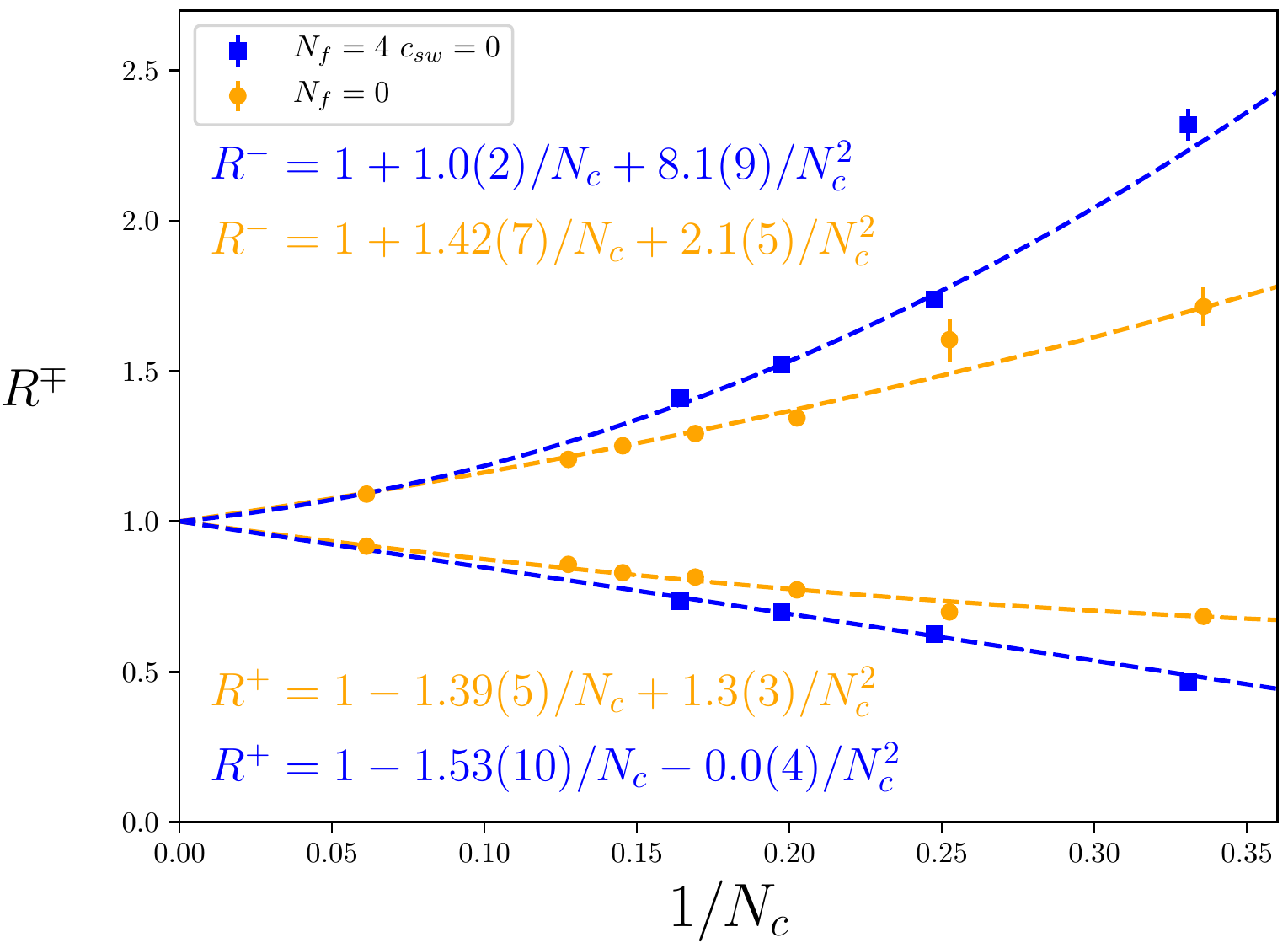}
\caption{ $K \to \pi$ matrix elements as a function of the number of colors for dynamical and quenched configurations. The label $c_{sw}=0$ in the legend refers to the value of $c_{sw}$ in the valence action.    \label{fig:ktopi} }
\end{figure}

We show our preliminary results in Fig. \ref{fig:ktopi} for the dynamical ensembles of this work, together with the quenched results of Ref. \cite{Donini:2016lwz}. We also fit these numbers to the expected $N_c$ dependence. We obtain that the $1/N_c$ terms agree in both calculations, and that quenching effects are sizeable and enter at $O(N_c^{-2})$. With these results, we can calculate the ratio of isospin amplitudes for $N_c=3$ using Eq. (\ref{eq:ratiog}):
\begin{equation}
\frac{A_0}{A_2}\Big|_{N_c=3,N_f=4} = 6.3(3), \ \ \ \ \frac{A_0}{A_2}\Big|_{N_c=3,N_f=0} = 3.7(1),
\end{equation}
where the error is purely statistical. In addition, an extra $10-20\%$ effect can be explained from a contribution to NLO ChPT (See Ref. \cite{Romero-Lopez:2018rzy}). Still, a factor three remains to be explained. 

\section{Conclusion and Outlook}

We have presented the current status of the determination of the $N_c$ scaling of QCD observables from first principles. We have achieved significant progress in the computation of meson masses and decay constant, the $I=2$ $\pi \pi$ scattering length and $K \to \pi$ matrix elements. 
First, we have been able to extract the leading and subleading contributions to the LECs of the chiral  Lagrangian. Second, we have seen the expected scaling in the $I=2$ $\pi \pi$ scattering length. Finally, the study of the $K \to \pi$ matrix elements has shown that $1/N_c^2$ effects are relevant and significantly increase the ratio of couplings. This is related to the $\Delta I = 1/2$ rule, and constitutes a source of enhancement.

In future work, we intend to address the scaling of QCD resonances. In particular,  $I=0$ $\pi \pi$ scattering length represents a very interesting topic, as it couples to the sigma resonance and it could have implications for the $\Delta I = 1/2$ rule. Other relevant topics are the $\rho$ resonance, the flavour singlet meson and the existence of tetraquarks in the large $N_c$ limit.

\section*{Acknowledments}
FRL, AD and PH acknowledge the support of the European Projects H2020-MSCA-RISE-2015/690575-InvisiblesPlus and H2020-MSCA-ITN-2015/674896-ELUSIVES. FRL, AD and PH have also received funding through the projects FPA2017-85985-P, SEV-2014-0398 and PROMETEO/2019/083. Moreover, the work of FRL has received funding from the European Union Horizon 2020 research and innovation program under the Marie Skłodowska-Curie grant agreement No. 713673 and "La Caixa" Foundation (ID 100010434, LCF/BQ/IN17/11620044). CP thankfully acknowledges support through the Spanish projects FPA2015-68541-P (MINECO/FEDER) and PGC2018- 094857-B-I00, the Centro de Excelencia Severo Ochoa Programme SEV-2016-0597, and the EU H2020-MSCA- ITN-2018-813942 (EuroPLEx). We thank Mare Nostrum 4 (BSC), Finis Terrae II (CESGA), Cal\'endula (SCAYLE), Tirant 3 (UV) and Lluis Vives (Servei d'Informàtica UV) for the computing time provided.

\end{document}